# GDF - A GENERAL DATAFORMAT FOR BIOSIGNALS VERSION V2.5123

Alois Schlögl[1,2]
<alois.schloegl@ist.ac.at>

[1]Institute for Human-Computer Interfaces, Graz University of Technology, Austria.
[2]Institute of Science and Technology, Klosterneuburg, Austria

ToDo:
- define StartTime in UTC or localtime
- describe usage of sparse sampling
- EVE NT.{POS,DUR} 64bit ?

Changes with v2.22
+ TOffset added
Changes with v2.23
+ Change references from http://...tu-graz.ac.at/ to http://pub.ist.ac.at/
Changes with v2.51
+ add TimeStamps in event table
+ Time zone information is mandatory
+ SCP-ECG Section 7-11 can be stored in Header 3 (tags 9-13)

**ABSTRACT**
Biomedical signals are stored in many different data formats. Most formats have been developed for a specific purpose of a specialized community for ECG research, EEG analysis, sleep research, etc. So far none of the existing formats can be considered a general purpose data format for biomedical signals. In order to solve this problem and to unify the various needs of the various biomedical signal processing fields, the so-called General Data Format for biomedical signals" (GDF) is developed. This GDF format is fully described and specified. Software for reading and writing GDF data is implemented in Octave/Matlab and C/C++ and provided through BioSig – a free and open source software library for biomedical signal processing. BioSig provides also converters from various data formats to GDF, and a viewing and scoring software.

# 1. INTRODUCTION

Biosignals are currently stored in a wide variety of different data-formats. BioSig – an open source software library – supports approximately 50 different data formats [1]. Many data formats have been developed by commercial companies, research groups and standardization organizations. Furthermore, some scientific data formats like HDF, netCDF and CDF should be mentioned, however these are not sufficient to store biomedical signals in an unambiguous manner. An overview is provided at [2,21]. The BioSig project [1] supports more data formats (currently ca. 40) than any other software project in the field, and provides also converters for a number of data formats [16]. A detailed comparison of about 20 data formats for biomedical signals (those with an public available specification) is available, too [20].

Most of the data formats addressed just some specific needs, but none was flexible enough to address new needs in the field of EEG recordings and Brain-Computer Interface research. Therefore, version 1 of the General Data format (GDF) [3] has been developed, and was successfully implemented and used in BCI research. It provides a common coding scheme for events, and supports many useful features (different sampling rates and calibration values for different channels, an automated overflow detection, support of different data types, encoding of filter settings etc.), that are only partly implemented in other formats. Within the recent years, new requirements became apparent. In this report, Version 2 of the GDF is described. Briefly, it addresses the need for
- Subject specific information (gender, age, impairment, etc)
- Recording location, identification of recording software, etc.
- Possibilities for storing the electrode positions in spatial coordinates, electrode impedances, etc.
- More efficient encoding of date and time, physical dimensions, filter information
- Non-equidistant (sparse) sampling

The structure of GDF v2.0 is similar to EDF [4], GDF1.x [3] and EDF+ [5], all of them have a fixed header with 256 bytes and a variable header with 256 times "number_of_channels". Moreover, GDF v2.0 supports storing events (annotations and markers) and non-equidistant (sparsely) sampled data. Based in the version identification field (first 8 bytes), it is easy to write software supporting all formats simultaneously. This is also demonstrated by the BioSig project [1], which provides the prototype implementation for Octave/Matlab as well as C/C++. Support for other programming languages (like Python and Java) is under consideration.

With GDF v2.1 the (free header) has been changed to a structured header (Header 3). The structure of header3 is based on a Tag-Length-Value structure and contains optional fields, not defined in any of the other fields. Each tag may occur at most once. Currently, the description of the user-specified description of event codes (tag=1) and information related to the BCI2000 system (tag=2) are defined; tag=255 can be used for arbitrary information (former free header). All other flags are reserved for future use (Table 10).

# 2. SPECIFICATION

The general data structure is defined in Table 9 (see Annex). An GDF file consists of the following five components: the fixed header or header 1 (with 256 bytes) is mandatory, the variable header or header 2 containing channel-specific information (number-of-channels times 256 bytes), the TLV header or header 3 contains optional information, the data section, and the table of events. Header 2 can be empty, in case that no channel information is stored (e.g. in pure event files).

*Table 1: Structure of a GDF file. NS, NT, and NEV indicate the number of channels, number of TLV elements, and the number of events, respectively,*

|   | Description | size [bytes] |   |
|---|---|---|---|
| 1 | Fixed header, (header1) | 256 | 1 |
| 2 | variable header, channel-specific information (header 2) | 256*NS | 1 |
| 3 | optional header, tag-length-value structure, (header 3) | >=4*NT + length of each value field <=Length of header – 256*(NS+1) | 0 <= NT <256 |
| 4 | data section | number_of_records times bytes per record | <=1 |
| 5 | event table | 8 + NEV*6 {mode 1} or <br> 8 + NEV*12 {mode 3} | <=1 |

The sections are described in more detail in Table 9 (see Annex). The issues were considered during the design of the data format.

1. Data is stored in little endian format. However, BioSig [1] supports also big-endian platforms by converting the data internally.
2. The Version field is of type char[8] and is stored at the beginning of the file. The present specification requires "GDF 2.10". This field is used to provide upwards compatibility with past and future versions of GDF.

*Table 2: Encoding of 85$^{th}$ byte.*

| Byte 85 | bit 7 | bit 6 | bit 5 | bit 4 | bit 3 | bit 2 | bit 1 | bit 0 |
|---|---|---|---|---|---|---|---|---|
| **Smoking** | | | | | | | | |
| unknown | | | | | | | 0 | 0 |
| NO | | | | | | | 0 | 1 |
| YES | | | | | | | 1 | 0 |
| | | | | | | | | |
| **Alcohol abuse** | | | | | | | | |
| unknown | | | | | | 0 | 0 | |
| NO | | | | | | 0 | 1 | |
| YES | | | | | | 1 | 0 | |
| | | | | | | | | |
| **Drug abuse** | | | | | | | | |
| unknown | | | | 0 | 0 | | | |
| NO | | | | 0 | 1 | | | |
| YES | | | | 1 | 0 | | | |
| | | | | | | | | |
| **Medication** | | | | | | | | |
| unknown | | 0 | 0 | | | | | |
| NO | | 0 | 1 | | | | | |
| YES | | 1 | 0 | | | | | |
| | | | | | | | | |

3. The patient identification field is 66 bytes long (instead of 80 bytes in EDF [4], EDF+ [5] and GDF 1.x [3]). This fact is justified, because in EDF and GDF 1.x rarely all 80 bytes are used. In EDF+ [5] 14 bytes of this field are used for encoding Gender and Birthday; GDF 2.0 stores this information elsewhere (the gender is encoded in byte 88 and the Birthday is binary encoded in bytes 177-184). Therefore, the PID field is reduced by 14 bytes. The remaining 66 bytes contain the patient identification code, the patient name, the patient classification. The use of any remaining bytes is not specified. Each field is separated by and must not contain a space (ASCII 32) character. Empty fields are encoded by the single letter "X"; For reasons of privacy protection, we recommend not to include the real patient name.

*Table 3: Encoding of 88$^{th}$ byte*

| Byte 88 | bit 5 | bit 4 | bit 5 | bit 4 | bit 3 | bit 2 | bit 1 | bit 0 |
|---|---|---|---|---|---|---|---|---|
| **Gender** | | | | | | | | |
| unknown | | | | | | | 0 | 0 |
| male | | | | | | | 0 | 1 |
| Female | | | | | | | 1 | 0 |
| unspecified | | | | | | | 1 | 1 |
| **Handedness** | | | | | | | | |
| Unknown/undefined | | | | | 0 | 0 | | |
| Right | | | | | 0 | 1 | | |
| left | | | | | 1 | 0 | | |
| equal | | | | | 1 | 1 | | |
| | | | | | | | | |
| **Visual inpairment** | | | | | | | | |
| Unknown/undefined | | | 0 | 0 | | | | |
| No impairment | | | 0 | 1 | | | | |
| Visual impairment | | | 1 | 0 | | | | |
| Impairment but Corrected | | | 1 | 1 | | | | |
| | | | | | | | | |
| **Heart impairment** | | | | | | | | |
| Unknown | 0 | 0 | | | | | | |
| no | 0 | 1 | | | | | | |
| yes | 1 | 0 | | | | | | |
| pacemaker | 1 | 1 | | | | | | |

4. The 85$^{th}$ byte (see Table 2) can be used to describe smoking, alcohol abuse, drug abuse and/or medication.
5. The weight field contains the body weight in kilograms [kg]; a value of 0 indicates an unknown weight, a value of 255 indicates a weight larger then 254 kg.
6. The height field contains the body height in centimeters [cm]; a value of 0 indicates an unknown height, a value of 255 indicates a height larger then 254 cm.
7. The gender, handedness and visual impairment of the subject is stored the 88$^{th}$ byte according to Table 3.

8. The field recording identification (RID, starting at byte 89) contains an identification for the investigation or scientific study. In addition a consecutive number can be included. Compared to EDF this field has been reduced from a length of 80 to 68 bytes. This is justified because EDF+ uses 12 bytes to store the recording date; in GDF this information is stored in bytes 169-172. If byte 156 is zero, bytes 153-156 are used by the RFC1876 header (see next issue) and must not be used by the RID field.
9. The recording location (place of recording) is stored according the RFC1876 using 4 uint32 numbers (bytes 153 to 168 in the GDF header, see Table 4). If the version number (byte 156) is not zero, bytes 153-156 are used by the field "recording identification" and do not belong to the location information.

*Table 4: Encoding of the recording location*

| GDF Header bytes | |
|---|---|
| 153 | Vertical Precision, byte |
| 154 | Horizontal Precision, byte |
| 155 | Size, byte |
| 156 | Version, byte |
| 157-160 | Latitude [1/3600000 degrees], int32 |
| 161-164 | Longitude [1/3600000 degrees], int32 |
| 165-168 | Altitude [in cm], int32 |

10. The date/time format of the begin of data recording is changed in Version 2.0 from an 16 byte ASCII format to an 8-byte (64 bit) binary format. The 4 least significant bytes indicate the fraction of a day ($2^{32}$ corresponds to $3\,600 * 24 = 86\,400s$ or 1 day) and provides a time resolution of approx 20 µs. The 4 most significant bytes count the number of the days since 1. The time is stored as local time, v2.5 or later supports storing the time zone information (see below). Jan 0000. Jan 1st, 1970 is day number 719 529. Conversion of this integer into a double (with 52 bits mantissa) does not result in round-off errors for the next 800 years. Only in the year 2800, the last bit will be lost which results in a time resolution of 40µs instead of 20µs for another 2800 years. Accordingly, this date/time format sufficiently accurate for storing the starting time and birthday information. The starting time of the recording is stored n bytes 169-176, the birthday of the subject is stored in bytes 177-184 for obtaining the subjects age at recording time. However, for privacy protection, the birthday could have been modified up to 1 year. The date-time conversion can be implemented in the following way:

```
/* conversions between t_time and gdf_time */

gdf_time = (int64) round((t_time/(3600*24) + 719529) * 2^32);
t_time = ((gdftime / 2.0^32) – 719529) * (3600*24);
```

11. The length of the header is defined in the field "number of blocks (each 256 bytes) in the header record". The number of blocks must be at least (1+NS). The number of data blocks beyond (NS+1) are used for header 3.
12. Equipment Provider Identification is is defined as 8 byte code. Application for codes should be addressed to the author.
13. Technician ID and Lab ID have been moved to header 3 (tag 6 and 7). The IP address of the recording equipment is also included in header 3 (tag 5).
14. Starting with version 2.1, bytes 187-192 are used for patient classification according to the International Statistical Classification of Diseases and Related Health Problems (ICD). ICD9 or ICD10 can be used. Patient classification according to Snomed CT can be stored in header 3, tag 8. The size of the subject's head (circumference, distance nasion – inion, left to right mastoid) is stored in three 16-bit integer numbers in the bytes 207-212. The distances are stored in the units millimeter [mm]. Value 0 indicates an unknown value.

Table 5: Examples of Physical Units. The full table is described in Annex A of the FEF standard and is available from biosig4matlab/doc/units.csv http://biosig.svn.sourceforge.net/viewvc/biosig/trunk/biosig/doc/units.csv [1, 11(p. 62-75, Table A.6.3), 12].

| Code | Units | Description |
|---|---|---|
| 0 | Unknown/undefined | |
| 512 | - | Dimensionless |
| 544 | % | Per cent |
| 736 | degree | |
| 768 | rad | |
| 2496 | Hz | |
| 3872 | mmHg | Blood pressure |
| 4256 | V | Voltage |
| 4288 | Ohm | Resistance, impedance |
| 4384 | K | Temperature in Kelvin |
| 6048 | °C | Temperature in degree Celsius |
| | | |
| 3072 | l/min | liter per minute |
| 2848 | l/(min m^2) | liter per minute square meter |
| 4128 | dyn s / cm^5 | hydraulic impedance |
| 6016 | dyn s / m^2 cm^5 | Pulmonary/ Systemic Vascular Resistance Index |
| | | |

Table 6: Table of Decimal Factors. The table is also available from biosig/doc/DecimalFactors.txt http://biosig.svn.sourceforge.net/viewvc/biosig/trunk/biosig/doc/DecimalFactors.txt [1, 11 (p. 60, Table A.6.1), 12].

| Name | Magnitude | Code Offset |
|---|---|---|
| yotta | $10^{+24}$ | 10 |
| zetta | $10^{+21}$ | 9 |
| exa | $10^{+18}$ | 8 |
| peta | $10^{+15}$ | 7 |
| tera | $10^{+12}$ | 6 |
| giga | $10^{+9}$ | 5 |
| mega | $10^{+6}$ | 4 |
| kilo | $10^{+3}$ | 3 |
| hecto | $10^{+2}$ | 2 |
| deca | $10^{+1}$ | 1 |
| | $10^{+0}$ | 0 |
| deci | $10^{-1}$ | 16 |
| centi | $10^{-2}$ | 17 |
| milli | $10^{-3}$ | 18 |
| micro | $10^{-6}$ | 19 |
| nano | $10^{-9}$ | 20 |
| pico | $10^{-12}$ | 21 |
| femto | $10^{-15}$ | 22 |
| atto | $10^{-18}$ | 23 |

| zepto | 10-21 | 24 |
| yocto | 10-24 | 25 |

15. A "record" is a data segment with a fixed length as indicated in the field "Duration". The number of data records is of type int64 (8byte); the duration of data records in seconds is a rational number consisting of the numerator (uint32) and a denominator (uint32) (together 8byte). The number of channels is of type uint32 (4byte).
16. The field "NS" indicates the number of channels.
17. The "label"-field contains the description for each channel. It is recommended that the labels follow existing standards (e.g. [11,12]) whenever possible. The transducer type describes the sensor type. There exist already some standards for encoding of the channels and transducers (e.g. EN1064 [10] and [11,12]). Currently, the encoding scheme is not complete. Therefore, these fields are ASCII-encoded. However, this might change in future.
18. The Physical Dimensions are encoded with a 16 bit integer number according to The Annex of the File Exchange Format (FEF) for Vital signs [11,12]. The Physical Dimensions are encoded according to [11] (Table A.6.1, p.52-53, Table A.6.3, p.54-61) and [12] (Table A.4.3 p. 38- 44). Some examples are listed in Table 5. The decimal factors are encoded as offsets (using the 5 least significant bits) according to Table 6. For example, the physical dimension "µV" (micro-Volt) is encoded as 4256 (Volt) + 19 (micro) = 4275.
19. Physical minimum and maximum are stored as IEEE floating point numbers with double precision (8 bytes) instead of char[8] (=8*ASCII).

*Table 7: Data types, memory requirement, coding scheme. Similar data types are defined in [13].*

| data type (GDFTYP) | size per sample [bytes] | code | Range [min .. max], remarks |
|---|---|---|---|
| int8, INT-I8 | 1 | 1 | -128 ..127 |
| uint8, INT-U8 | 1 | 2 | 0 .. 255 |
| int16, INT-I16 | 2 | 3 | -32 768 .. 32 767 |
| uint16, INT-U16 | 2 | 4 | 0 .. 65 535 |
| int32, INT-I32 | 4 | 5 | -2 147 483 648 .. 2 147 483 647 |
| uint32, INT-U32 | 4 | 6 | 0 .. 4 294 967 295 |
| int64 | 8 | 7 | -2^63 .. 2^63-1 |
| uint64 | 8 | 8 | 0 .. 2^64-1 |
| | | | |
| float32, | 4 | 16 | According to [14] |
| float64 | 8 | 17 | According to [14] |
| float128 | 16 | 18 | According to [15] |
| | | | |
| int24, INT-I24 | 3 | 279 | -8 388 608 .. 8 388 607 |
| | | | |
| uint24, INT-U24 | 3 | 535 | 0 .. 16 777 215 |
| | | | |

20. Digital minimum (DigMin) and maximum (DigMax) are of type "float64" and shall not exceed the range of the data type (see Table 7). This field has 52 bits significants, therefore all integer numbers up to 52 bits are stored in its full precision. In order to support quality control and support an automated overflow detection [6], the value of digital minimum and maximum should indicate the overflow (saturation) values of the recording system. Moreover, a sample value not within the interval between DigMin and DigMax indicates an invalid measurement

20. (e.g. caused by an over- or underflow, or when the sensor is off in intermittently sampled data). If there is a need to distinguish between overflow, underflow and sensor-off, distinct values should be used for each condition.
21. The field "pre-filtering" is in GDF Version 2.0 only 68 bytes (it was 80 bytes in EDF, BDF and GDF Version <1.9) and will become obsolete in future. Its purpose is replaced by the filter settings.
22. The filter settings are stored as floating point numbers in the variable header. Low pass, High pass and Notch are stored as a 32-bit floating point number. Unknown values are indicated by NaN's; Notch Off is indicated by a number smaller than zero (typically -1).
23. The "Samples Per Record"-field is uint32 (4bytes)*NS. The "Samples Per Record" is the sampling rate of the corresponding channel multiplied by the duration of a record. A value of zero (samples per record), indicates a channel with sparse (non-equidistant) sampling. In this case, the actual sampling values are stored in the event table.
24. Each channel can use a different data type . The type information (see Table 7 ) of each channel is stored in the variable header after "number of samples per record".

25. The X-Y-Z coordinates of the EEG electrodes are included in the variable header. The coordinates of the references and ground electrodes are included in the fixed header from bytes 213-224 and 225-236, respectively. All positions should be stored in a coordinate system according to [9].
26. With GDF version 2.19, the last 20*NS in the variable header 2 (bytes 256+236*NS .. 256+256*NS) are redefined. For version < 2.19 the impedance of the electrode will be stored in compressed form in 1 byte. For this purpose, the digital value will correspond to the impedance according to the following formulas.

    ```
    DIGVAL = round(log2(Z[Ohm])* 8)
    Z[Ohm] = 2^(DIGVAL/8)
    ```

    This allows to store the value of the electrode impedance with less than +-5 % relative error. DIGVAL is 255, if the impedance is undefined or larger than $2^{(254/8)}$=3.9 GOhm.

    With version 2.19 and later, this area is used for context- and channel specific information (see Table 8). If the channel stores voltage data, (physical unit with base code 4256), the impedance of the electrode will be stored in float32 number. If the channel contains impedance values (physical unit "Ohm, physical dimension with base code 4288), the probe frequency is stored as a float32 number. Unknown values should use NaN (not-a-number). The unused bytes of header2 are reserved for future use.

*Table 8: Sensor-specific information. Depending on the data/transducer/sensor, the following information is added in the sensor-specific field (last NS*20 bytes in header 2). The context is determined by the physical dimension (see Table 5).*

| Type of data | PhysDimCode & 0xFFE0 | Sensor-specific information | | Reserved |
|---|---|---|---|---|
| Voltage (EEG, ECG, etc.) | 4256 | Electrode impedance [Ohm] | float32 | 16 bytes |
| Impedance | 4288 | Probe frequency [Hertz] | float32 | 16 bytes |
| others | | - | - | 20 bytes |

27. The field "TimeOffset"(TOffset) indicates the time delay between channels in relation to some

references channel. If all channels are sampled in parallel, (i.e. no time delay), this value is zero. Otherwise it indicates the delay of the sample times between the channels. This value always smaller than the smallest sampling interval. TOffset is a floating point number indicating the time delay in seconds. If the time delay is not known, the value 'not-a-number' (NaN) should be used.

28. With version 2.10, Header 3 contains a list of tag-length-value elements. Free header information as allowed in GDF 2.0 and earlier can be stored within the element with tag=255. The maximum size of header 3 is HeadLen – (NS+1)*256. Because the header length is a multiple of 256 bytes, there might be up to 255 unused bytes. Unused bytes must be set to 0. The list of TLV-elements is terminated when an element with tag=0 is found, or when there are less then 4 bytes (to read tag and length) for the next element.
29. The data section contains the samples of the equidistant sampled channels. The samples of each channel and record are stored consecutively. First all samples of the first channel from the first record are stored, then all samples of channel 2 of the first block, until all channels of the first block are stored. Then, the samples of the second block are stored in the same order.
30. The table of events is stored after the data section. The starting position of the event table (event table position ETP) can be calculated in the following way.

    ETP = Length_of_header + Number_of_Records * Bytes_per_Record.

    The value of the first byte (mode of event table) can be 1 or 3. A value of 1 indicates that the event table contains event-type and position. A value of 3 indicates that the position, type, associated channel and the duration of the event is stored.
31. Bytes 2 to 4 of the event table represent a 24-bit integer value indicating the number of events.
32. The next four bytes contain a floating point number representing the sample rate associated with the event position. Dividing the event position (and duration) by this value yields the position (duration) in seconds.
33. Then the position of all events is saved in 32-bit integers using a one-based indexing (position of first sample is 1, not 0), followed by the event type as 16-bit integers. The encoding of the various event types is defined in Table 11 (see Annex). If the mode of the event table has value 1, each event is defined by the event type and position {TYP,POS}. In case the mode of the event table is 3, the channel number associated with each event (a value of 0 indicates the event refers to all channels) and the duration of the event are included too {TYP,POS,DUR,CHN}. If the channel number refers to a non-equidistant (sparsely) sampled channel and the event type is 0x7fff, instead of the duration the value of a the sparsely sampled channel is represented. The scaling (Physical and digital minimum and maximum) applies to the sample value.
34. The information about the recording device is stored in the header 3 (tag=3) contains the name of the manufacturer, the model name, the model version, and the serial number. Each of these for fields are a zero-terminated string. If some field is not used, only the terminating zero is stored.
35. Some sensor (like MEG) have not only a position but require also an orientation for its full characterization. The orientation of such sensors can be stored (header3, tag=4) as an 3 x NS array of float32 [IEEE754].
36. The time zone information can be stored in bytes 254 and 255 as int16 number starting from Version 2.40. This field is mandatory for V2.50 and later. The time zone information is stored as time offset in minutes to UTC, e.g. for Berlin, its +60 in winter and +120 in summer. It applies to the startdate/time as well as to the timestamp information in the event table. The starttime and all timestamps are stored in the GDF representation of its localtime. This is comparable to SCP-ECG (see [10], section 1, tag 34).
37. In order to provide a lossless conversion of data in SCP format [10],. GDF v2.50 and later can store the SCP Section 7-11 in the optional GDF Header 3 (tags 9-13)(see Table 10). The SCP sections are stored without its 16 byte "Section ID Header". The detailed specification of these sections is described in [10].
38.

## 3. RECOMMENDATIONS AND REMARKS:

GDF v.1 defined also data types of bit lengths e.g. bit 1, bit7, bit 4 etc. However, this has not been much of use. Furthermore, it complicated software development. Therefore, the use of bit-types is discouraged in GDF v.2.

The event table is defined only, if the number of records is known. Consequently, no event table can be stored in an ongoing recording, because the number of records is usually not known (NRec = -1). The event table can be only written to the GDF file, once the recording length (i.e. data size, number of records) is known (often only after the end of the data acquisition).

The begin and end of an event can be stored in two different ways, either in {TYP, POS} (event table mode=1) or with {TYP,POS,DUR,CHN}. In the former case, the begin and the end is stored in two separate event entries, while the end (offset) of the event is encoded with the highest bit set, i.e. or-ing the type of the start event with 0x8000.In the latter case, the end of the event is at the position POS+DUR.

Non-equidistant sampled data can be stored, too. For this purpose, an additional channel header must be defined, and the corresponding sampling rate must be set to zero. The actual samples are stored in the event table, and the channel information must contain the corresponding channel number, then the 32 bit duration field is used for the actual sample value. T he data type must correspond to the value of GDFTYP of the respective channel, and must not contain more than 32 bits. Preferably, the data type should be of type uint32. Intermittently sampled data (on/off periods) can be stored either as non-equidistant sampled data (described above), or as fully sampled data where the values is not within the interval between DigMin and DigMax when the sensor is off.
The format definition of GDF is nearly as simple as the definition of EDF [4], as can be seen in Table 9. The use of binary encodings enables a more compressed representation; accordingly, more information can be stored within the header information. This enables a higher accuracy (e.g. in date and time information) and additional information can be stored without extending the header size.

The proposed format specification was successfully implemented in C/C++ as well as an M-file which can be used with Octave (>2.9.12) and Matlab® (>6.5). The software implementation requires only minor changes to upgrade from EDF, BDF or earlier GDF to GDF 2.2x. The software is available "online" (see BioSig – an open source software library for biomedical signal processing [1]) and is "free" under the terms of the "General Public License" (GPL) [7]. BioSig demonstrates also how different data formats can be simultaneously supported. This demonstrated how a smooth upgrade to the more feature-rich GDF file format can be implemented, without breaking backwards compatibility and without additional workload for the user.

## 4. DISCUSSION AND CONCLUSION

It is reasonable to say that GDF provides a superset of features from many other data formats. GDF v2.10 includes support for, user-specified event description (like in EDF+ and BrainVision format), manufacture information (like in SCP [10] and MFER [19]), and the orientation of MEG are supported. Accordingly, GDF v2.x is (upwards) compatible to most other data formats; this means biosignal data from other formats can be converted to GDF without loss of information. Routines for reading and writing of GDF files in Octave and Matlab as well as in C are implemented in the open source package BioSig [1].

## 5. ACKNOWLEDGMENT

I thank (in alphabethical order) Martin Billinger, Clemens Brunner, Oliver Filz, Arnulf Heller, Peter Jacobi, Claudia Keinrath, Owen Kelly, Jürgen Mellinger, Herbert Ramoser, Stefan Sauermann, Reinhold Scherer and Darren Weber for their contributions to the discussion.

## 6. REFERENCES

[1] A. Schlögl, The BioSig project. 2003-2013. online available at: http://biosig.sf.net/ .
[2] Overview of different data formats: http://pub.ist.ac.at/~schloegl/matlab/eeg/

[3] A. Schlögl, O. Filz, H. Ramoser, G. Pfurtscheller, GDF - A general dataformat for biosignals, Technical Report, 2004. available online at:
http://pub.ist.ac.at/~schloegl/matlab/eeg/gdf4/TR_GDF.pdf

[4] B. Kemp, A. Värri, A.C. Rosa, K.D. Nielsen and J. Gade (1992): A simple format for exchange of digitized polygraphic recordings. *Electroenceph. clin. Neurophysiol.,* 82: 391-393.

[5] Kemp B, Olivan J. European data format 'plus' (EDF+), an EDF alike standard format for the exchange of physiological data. *Clin Neurophysiol.* 2003 Sep;114(9):1755-61.

[6] A. Schlögl, B. Kemp, T. Penzel, D. Kunz, S.-L. Himanen, A. Värri, G. Dorffner, G. Pfurtscheller. Quality Control of polysomnographic Sleep Data by Histogram and Entropy Analysis. *Electroenceph. clin. Neurophysiol.* 1999, Dec; 110(12): 2165 - 2170.

[7] The General Purpose License http://www.fsf.org/

[8] RFC1876 How latitude and longitude are stored in a DNS record. available online: http://www.faqs.org/rfcs/rfc1876.html

[9] J.A. Malmivuo. Consistent System of Rectangular and Spherical Coordinates for Electrocardiology and Magnetocardiology. online available:
http://butler.cc.tut.fi/~malmivuo/bem/bembook/aa/aa.htm.

[10] ISO 11073-91064:2009 Health informatics -- Standard communication protocol -- Part 91064: Computer-assisted electrocardiography

SCP-ECG, Health informatics - Standard Communication Protocol - Computer-assisted electrocardiography, CEN TC 251/N02-015 prEN1064:2002. CEN/TC251 Secretariat: SIS - Swedish Standards Institute.

[11] ISO/IEEE 11073-10101:2004, Health informatics - point-of-care medical device communication Part 10101: nomenclature.

[12] CEN/TC251 PT40, File Exchange Format of Vital Signal Annex A (Normative).

[13] ISO/IEEE 11073-20101, Health informatics - point-of-care medical device communication Part 20101: Application profiles - Base standard. Appendix A, Table A.1. p. 19.

[14] IEEE Standard for Binary Floating-Point Arithmetic (ANSI/IEEE Std 754-1985), also known as IEC 60559:1989, Binary floating-point arithmetic for microprocessor systems.

[15] Revising ANSI/IEEE Std 754-1985, http://754r.ucbtest.org

[16] A. Schlögl, F. Chiarugi, E. Cervesato, E. Apostolopoulos, C. Chronaki.
Two-Way Converter between the HL7 aECG and SCP-ECG Data Formats Using BioSig.
Computers in Cardiology Conference. p253-256, 2007.

[17] RFC 791, Internet Protocol – DARPA Internet Protocol Specification. Available online http://tools.ietf.org/html/rfc791, 1981.

[18] RFC2460, Internet Protocol Version 6 (IPv6) Specification. Available online: http://tools.ietf.org/html/rfc2460, 1998.

[19] Medical waveform format (MFER) ISO/TS 11073/92001:2007.

[20] A. Schlögl. An overview on data formats for biomedical signals. - in: Image Processing, Biosignal Processing, Modelling and Simulation, Biomechanics (2009), p. 1557 – 1560 .
World Congress on Medical Physics and Biomedical Engineering, Munich 2009.
https://online.tu-graz.ac.at/tug_online/voe_main2.getVollText?pDocumentNr=111055&pCurrPk=45404

[21] http://en.wikipedia.org/wiki/List_of_file_formats#Biomedical_Signals_.28Time_Series.29

*Table 9: Overview of the header definition for GDF 2.0. The "remark" column indicates which fields have changed form Ver 1.25 to Ver 2.x.*

| FIXED HEADER (Header 1) | Remark 1.x -> 2.0 | Position Start:End [bytes] | Bytes | Type |
|---|---|---|---|---|
| Version identification (GDF 2.00) | updated | 0 | 8 | char[8] |
| Patient identification (P-id) | smaller | 8 | 66 | char[80] |
| reserved | | 74 | 10 | uint8 |
| Smoking / Alcohol abuse / drug abuse /medication (see Table 2) | Added | 84 | 1 | bit[4x2] |
| Weight [in kg] | Added | 85 | 1 | uint8 |
| Height [in cm] | Added | 86 | 1 | uint8 |
| Gender [bits 0..1] (see Table 3) | Added | 87 | 2/8 | bit[2] |
| Handedness [bits 2..3] | Added | 87 | 2/8 | bit[2] |
| Visual impairment [bits 4..5] | Added | 87 | 2/8 | bit[2] |
| Heart impairment [bits 6..7] | Added | 87 | 2/8 | bit[2] |
| Recording identification (study id + serial number ) | smaller | 88 | 64 | char[64] |
| Recording Location (Lat, Long, Alt) (see Table 4) | Added | 152 | 16 | uint32[4] |
| Startdate and time of recording | changed | 168 | 8 | uint32[2] |
| Birthday | Added | 176 | 8 | uint32[2] |
| Header Length (number of 256-byte blocks) | smaller | 184 | 2 | uint16 |
| Patient classification according to ICD | NEW | 186 | 6 | byte[6] |
| Equipment Provider identification (EP-id) | | 192 | 8 | uint64 |
| Technician ID | removed | | 8 | |
| Lab ID | removed | | 8 | |
| reserved | | 200 | 6 | byte[6] |
| Headsize [in mm] | Added | 206 | 6 | uint16[3] |
| Position Reference Electrode [X,Y,Z] | Added | 212 | 12 | float32[3] |
| Position Ground Electrode [X,Y,Z] | Added | 224 | 12 | float32[3] |
| reserved | removed | | 20 | byte |
| number of data records (-1 if unknown) | | 236 | 8 | int64 |
| Duration of a data record, as a rational number in seconds (first the numerator, secondly the denominator. | | 244 | 8 | uint32[2] |
| NS: number of signals (channels) | | 252 | 2 | uint16 |
| Reserved time zone information (offset in minutes east of UTC) | v2.50 or later | 254 | 2 | uint16 |
| | | | | |
| VARIABLE HEADER (Header 2) | | | | |
| Label | | 256 | NS*16 | char[16]*NS |
| Type of Transducer/Sensor | | 256+16*NS | NS*80 | char[80]*NS |
| Physical dimension | obsolete | 256+96*NS | NS*6 | char[8]*NS |
| Physical dimension code ([11] Table A6.3) (see Table 5 and 6) | NEW | 256+102*NS | NS*2 | uint16*NS |
| Physical minimum (PhysMin) | | 256+104*NS | NS*8 | float64*NS |
| Physical maximum (PhysMax) | | 256+112*NS | NS*8 | float64*NS |
| digital minimum (DigMin) | changed | 256+120*NS | NS*8 | float64*NS |
| digital maximum (DigMax) | changed | 256+128*NS | NS*8 | float64*NS |
| Pre-filtering | obsolete | 256+136*NS | NS*64 | char[64]*NS |
| Time Offset in seconds (relative sampling time delay between channels) | v2.22 and later | 256+200*NS | NS*4 | float32*NS |
| Lowpass | Added | 256+204*NS | NS*4 | float32*NS |
| Highpass | added | 256+208*NS | NS*4 | float32*NS |
| Notch | added | 256+212*NS | NS*4 | float32*NS |
| Samples Per Record (0 indicates non-equidistant sampling) | | 256+216*NS | NS*4 | uint32*NS |
| Type of data (gdftyp) (see Table 7) | | 256+220*NS | NS*4 | uint32*NS |
| Sensor Position XYZ | added | 256+224*NS | NS*12 | float32*NS*3 |
| Sensor specific information (see Table 8) | v2.19 and later | 256+236*NS | NS*20 | byte[20]*NS |
| Electrode Impedance | only v2.0 – 2.18 | 256+236*NS | NS*1 | uint8*NS |
| reserved | only v2.0 – 2.18 | 256+237*NS | NS*19 | char[32]*NS |
| | | | | |
| **OPTIONAL Header (Header 3) (see Table 10)** | | 256*(NS+1) | | |
| Tag | NEW | | 1 | uint8 |
| Length | NEW | | 3 | uint24 |
| Value | NEW | | Length | (*) depends on tag |
| | | | | |
| **DATA RECORD** | | | | |
| nr samples from channel [1] of TYPE[1] | | | | Type[1] |
| nr samples from channel [2] of TYPE[2] | | | | Type[2] |
| nr samples from channel [3] of TYPE[3] | | | | Type[3] |
| ... | | | | ... |
| nr samples from channel [NS] of TYPE[NS] | | | | Type[NS] |
| | | | | |
| **EVENT TABLE** | NEW | ETP | | |
| mode of event table can be 1, ~~or~~ 3, 5 or 7.~~.~~ | | ETP + 0 | 1 | uint8 |
| Number of events NEV | Changed | ETP + 1 | 3 | uint8[3] |
| Samplerate associated with Event positions. | Changed | ETP + 4 | 4 | float32 |
| Position [in samples] | | ETP + 8 | NEV*4 | uint32 |
| Type (see Table 11) | | ETP+ 8+NEV*4 | NEV*2 | uint16 |
| Channel [optional] | ~~Only i~~If bit 2 in mode is set~~3~~ (mode & 0x02) | ETP+ 8+NEV*6 | NEV*2 | uint16 |
| Duration [in samples, optional] or Value of non-equidistant sampling | ~~Only i~~If bit 2 in mode is set~~3~~(mode & 0x02) | ETP+ 8+NEV*8 | NEV*4 | uint32 float32(*) |
| Time stamps | If bit 3 in mode is set (mode & 0x04) | ETP + 8 + NEV * {6 or 12} depending on (mode & 0x02) | NEV*8 | gdft_time |

*Table 10: Possible tags in Header 3.*

| Tag | Type of value | Description |
|---|---|---|
| 0 | | terminating tag; indicates last element in list of T-L-V elements |
| 1 | **char | list of null-terminated strings for user-specified description of event codes; last entry is identified by an additional empty string (in other words, the list is terminated by two consecutive zeros \0\0) . |
| 2 | *char | BCI 2000 header information (null-terminated string) |
| 3 | **char | Manufacturer name, model, version, serial number, stored as 4 consecutive zero-terminated strings. The total length (including the terminating zeros) should not exceed 128 bytes. |
| 4 | float32[NS][3] | orientation of MEG channels. For each sensor, three consecutive float32 values represent a vector in x-y-z direction indicating the direction of the sensor. There are NS (i.e. number of channels) vectors, thus the field has NS*3 numbers and needs NS*12 bytes. |
| 5 | IP-address | IP-address of recording computer in big endian format (hi byte first) The length is either 4 bytes for IPv4 [17] or 16 bytes for IPv6 [18] |
| 6 | | Technician identification |
| 7 | | Hospital/laboratory/clinic identification |
| 8 | *void | Patient classification according to Snomed CT. |
| 9 | *void | SCP ECG section 7 (without Section ID Header) [10] |
| 10 | *void | SCP ECG section 8 (without Section ID Header) [10] |
| 11 | *void | SCP ECG section 9 (without Section ID Header) [10] |
| 12 | *void | SCP ECG section 10 (without Section ID Header) [10] |
| 13 | *void | SCP ECG section 11 (without Section ID Header) [10] |
| | | |
| 8-254 | | reserved for future use |
| 255 (0xff) | | user specified, experimental, free header |

*Table 11: Table of event codes. The most recent version of this table will be available from [1]*
*http://biosig.svn.sourceforge.net/viewvc/biosig/trunk/biosig/doc/eventcodes.txt*

```
### Table of event codes.
# This file is part of the biosig project http://biosig.sf.net/
# Copyright (C) 2004-2011 Alois Schloegl <alois.schloegl@ist.ac.at>
# $Id: eventcodes.txt,v 1.3 2004/06/17 17:08:54 schloegl Exp $
### table of event codes: lines starting with # are omitted
### add 0x8000 to indicate end of event
#
### 0x010_      EEG artifacts
0x0101  artifact:EOG
0x0102  artifact:ECG
0x0103  artifact:EMG/Muscle
0x0104  artifact:Movement
0x0105  artifact:Failing Electrode
0x0106  artifact:Sweat
0x0107  artifact:50/60 Hz mains interference
0x0108  artifact:breathing
0x0109  artifact:pulse
### 0x011_      EEG patterns
0x0111  eeg:Sleep spindles
0x0112  eeg:K-complexes
0x0113  eeg:Saw-tooth waves
### 0x03__      Trigger, cues, classlabels,
0x0300  Trigger, start of Trial  (unspecific)
0x0301  Left - cue onset (BCI experiment)
0x0302  Right - cue onset (BCI experiment)
0x0303  Foot - cue onset (BCI experiment)
0x0304  Tongue - cue onset (BCI experiment)
0x0306  Down - cue onset (BCI experiment)
0x030C  Up - cue onset (BCI experiment)
0x030D  Feedback (continuous) - onset (BCI experiment)
0x030E  Feedback (discrete) - onset (BCI experiment)
0x0311  Beep (accustic stimulus, BCI experiment)
0x0312  Cross on screen (BCI experiment)
0x03ff  Rejection of whole trial
### 0x040_      Sleep-related Respiratory Events
0x0401  Obstructive Apnea/Hypopnea Event (OAHE)
0x0402  Respiratory Effort Related Arousal (RERA)
0x0403  Central Apnea/Hypopnea Event (CAHE)
0x0404  Cheyne-Stokes Breathing (CSB)
0x0405  Sleep Hypoventilation
### 0x041_      Sleep stages according to Rechtschaffen&Kales
0x0410  Wake
0x0411  Stage 1
0x0412  Stage 2
0x0413  Stage 3
0x0414  Stage 4
0x0415  REM
### 0x050_      ECG events
0x0501  ecg:Fiducial point of QRS complex
0x0502  ecg:P-wave
0x0503  ecg:Q-point
0x0504  ecg:R-point
0x0505  ecg:S-point
0x0506  ecg:T-point
0x0507  ecg:U-wave
### 0x____      OTHER
0x0000  No event
0x7FFF  non-equidistant sampled value
```